\documentclass[aps,prb,twocolumn,superscriptaddress]{revtex4-1}

\usepackage{amsfonts,amsmath,amssymb,mathbbol,mathtools}
\DeclareSymbolFontAlphabet{\mathbb}{AMSb}
\DeclareSymbolFontAlphabet{\mathbbl}{bbold}
\usepackage{array}
\usepackage[english]{babel}
\usepackage{color}
\usepackage{float}
\usepackage{xcolor}
\usepackage[normalem]{ulem}

\newcommand{\beq}{\begin{equation}}
\newcommand{\eeq}{\end{equation}}
\newcommand{\bea}{\begin{eqnarray}}
\newcommand{\eea}{\end{eqnarray}}
\def\br{{\mathbf{r}}}

\def\half{\tfrac{1}{2}}

\begin{document}

\title{Some Problems in Density Functional Theory}

\author{Jeffrey \surname{Wrighton}}
\email{jeffwrighton@gmail.com}
\affiliation{Department of Physics, University of Florida, Gainesville, Florida 32611}
\author{Angel \surname{Albavera-Mata}}
\affiliation{Center for Molecular Magnetic Quantum Materials, Quantum Theory Project, Department of Materials Science and Engineering, University of Florida, Gainesville, Florida 32611}
\author{H{\'e}ctor Francisco \surname{Rodr{\'i}guez}}
\affiliation{Department of Physics, Quantum Theory Project, University of Florida, Gainesville, Florida 32611}
\author{Tun S. \surname{Tan}}
\affiliation{Department of Physics, Quantum Theory Project, University of Florida, Gainesville, Florida 32611}
\author{Antonio C. \surname{Cancio}}
\affiliation{Department of Physics and Astronomy, Center for Computational Nanoscience, Ball State University, Muncie, Indiana, 47306, USA}
\author{J. W. \surname{Dufty}}
\email{dufty@ufl.edu}
\affiliation{Department of Physics, University of Florida, Gainesville, Florida 32611}
\author{S.B. \surname{Trickey}}
\email{trickey@ufl.edu}
\affiliation{Center for Molecular Magnetic Quantum Materials, Quantum Theory Project, Dept. of Physics, University of Florida, Gainesville, Florida 32611}

\date{24 June 2022; revised 18 Jan. 2023}

\begin{abstract}
Though calculations based on density functional theory (DFT) are used
remarkably widely in chemistry, physics, materials science, and
biomolecular research and though the modern form of DFT has been
studied for almost 60 years, some mathematical problems remain. From a
physical science perspective, it is far from clear whether those
problems are of major import.  For context, we provide an outline of
the basic structure of DFT as it is presented and used conventionally
in physical sciences, note some unresolved mathematical difficulties
with those conventional demonstrations, then pose several questions
regarding both the time-independent and time-dependent forms of DFT
that could benefit from attention in applied mathematics. Progress on
any of these would aid in development of better approximate
functionals and in interpretation of DFT.
\end{abstract}

\maketitle

%%%%%%%%%%%%%%%%%%%%%%%%%%%%%%%%%%%%%%%%%%%%%%%%%%%%%%%%%%%%%%%%%%%%%%%%%%%%%%%%%%%%%%%%%%%%%%%%%%%%
\section{Introduction and overview \label{sec:introduction}}
%%%%%%%%%%%%%%%%%%%%%%%%%%%%%%%%%%%%%%%%%%%%%%%%%%%%%%%%%%%%%%%%%%%%%%%%%%%%%%%%%%%%%%%%%%%%%%%%%%%%
Our objective is to identify some outstanding questions relating to
density functional theory (DFT),
including finite-temperature, orbital-free, and
time-dependent DFT.  

DFT is a reformulation of the Hamiltonian
description of quantum mechanics (e.g.\ Schr\"odinger equation) for
calculation of properties of interest (e.g. ground-state energy,
free-energy, etc.) As such, it is not an independent many-body theory,
hence adds no physics. Instead, it offers a framework that is quite
distinct from other approaches to the quantum mechanical many-body
problem.

Some readers may not be familiar with the extensive DFT literature of
the past sixty years. Therefore, we have organized our presentations regarding both
time-independent and time-dependent DFT to begin with contextual
overviews, Sections \ref{sec:ConvDFT} and \ref{sec:ConvTDDFT} respectively.
A generic but more abstract overview of the two is offered in the
Appendix \ref{sec:Appendix}.

By choice, our perspective for the two more physical overviews is the
conventional one of the physical sciences (condensed matter physics,
materials science, theoretical chemistry). In addition to being a fair
representation of our expertise (and its limits), the conventional
perspectives give a straightforward way to summarize known mathematical
difficulties with those approaches (a consequence, among
other things, of the different standards of proof typical of physical
sciences compared to mathematics).  Useful resources include
Refs. \onlinecite{EngelDreizlerBook,EschrigBook,KryachkoLudenaBook,EnglischEnglisch1984I,EnglischEnglisch1984II,ZumbachMaschke,Lieb83}.

Observe that our attention is restricted to quantum
systems. The counterpart DFT for classical systems, e.g. atoms in
liquid states \cite{Evans}, is an active area with problems of its
own. It is not discussed here.  Moreover, we restrict discussion to
many-electron systems because they, overwhelmingly, comprise the
quantum systems of interest in DFT.

Though our purpose is not a DFT review, it is important to note that 
DFT is, by now, the
most widely used many-electron method by far.  Its appeal and utility
rest in several aspects.  It reduces the many-degrees-of-freedom
problem of other many-fermion formulations to a functional dependency
on a single, physically observable scalar quantity, the electron
number density.  It provides a structure by which that density
(time-independent or time-dependent, zero or non-zero temperature) can
be generated by use of an auxiliary \textit{non-interacting}
many-electron system. That in turn, provides highly advantageous
computational cost-scaling with system size compared to other
many-body ans\"atze.  And it yields system-specific physical and
chemical property values (energies, free energies, ionization
potentials, \ldots etc.).  In detail, applications of DFT as a
computational tool confront many technical and fundamental
problems. Some of these are discussed in more detail below.

Before proceeding, some basic definitions are useful. We use Hartree
atomic units, $\hbar =m_{e}=q_{e}=1$ where $\hbar $ is Planck's
constant divided by $2\pi$, $m_{e}$ is the electron mass, and $q_{e}$
is the electron charge magnitude. We restrict consideration to
non-relativistic many-electron systems in an external potential such
that the system total energy (free-energy) per unit volume is bounded
below. The electron number density is $n(\mathbf{r})$; it is
normalized to the number of electrons (or number per cell as the case
may be), $N_{e}$.

The remainder of the paper is structured as follows.
Sec.~\ref{sec:ConvDFT} summarizes time-independent DFT as presented
conventionally along with the time-independent Kohn-Sham (KS) variational
procedure.  Sec.~\ref{sec:KnownDiffConvDFT} then gives the known
mathematical difficulties with those conventional approaches (both as
to the underlying theorems and the KS procedure), and poses four
questions about the scope and impact of those mathematical
deficiencies.  In Sec.~\ref{sec:OFdft}, we summarize the so-called orbital-free
formulation of KS DFT and pose six questions.  Section \ref{sec:OtherDFTissues}
poses six questions on various other aspects of time-independent DFT (e.g.
explicit number dependence, spin-density-functionals, spectral gap).  Turning
attention to time-dependent DFT (TDDFT), Sec.~\ref{sec:ConvTDDFT} summarizes conventional approaches
to formulating TDDFT.  Then Section \ref{sec:Issuestddft} %followed by
 discusses some fundamental issues of TDDFT proofs. 
We conclude with very brief remarks in
Sec.~\ref{sec:remarks}.

Throughout, we formulate specific questions that are of current interest to
the DFT community. About them, we believe the applied mathematics and
mathematical physics community may be able to make some helpful
progress.

\section{Time-independent DFT - conventional presentation \label{sec:ConvDFT}}

As noted above, time-independent DFT conventionally is presented as a
variational formalism within which the quantum mechanical energy of a
many-particle (usually many-fermion) system can be determined.  For a
system at zero temperature this refers to the ground state energy,
while at non-zero temperature (called ``finite temperature'' in the
physics literature) it refers to the thermodynamic free energy.

At the fundamental level the problem is characterized by a Hamiltonian $H$
for the system of interest plus an external potential coupling
separately to each degree of freedom $q_{i}$, $V=\sum v_{ext}(q_{i})$. The
energy is therefore a functional of this external potential, $E=E[v_{ext}].$
Similarly, derived properties also are functionals of $v_{ext}$.  For
example, the local number density $n(\br)$ is the functional derivative
$n(\br)=\delta E[v_{ext}]/\delta v_{ext}(\br)$. Therefore one may use
this to define
$n(\br\mid v_{ext})$ as a functional of $v_{ext}(\br)$. Via a Legendre transform a
functional of the density $F[n]$ can be constructed. In this way, the
properties of the system of interest are characterized by the density
for any external potential. The objective of time-independent DFT is
to find this representation through the propositions that (i) there is
a one-to-one relationship between $v_{ext}(\br)$ and $n(\br)$, and (ii) there
exists a universal (i.e. independent of $v_{ext}$) functional
$\mathcal{F}[n]$ such that the  extremum (minimum) of
$F[n] := \mathcal{F}[n] + \int d\br n(\br)v_{ext}(\br)$  occurs at $n_{0}(\br)$
with $F[n_0]$ being the ground state energy for temperature T=0
K or equilibrium free energy for T $>$ 0 K.  

Proofs of various forms of these propositions for many-electron
systems at zero temperature were given by Hohenberg and Kohn (HK)
\cite{HohenbergKohn} and refined by Levy \cite{Levy1979PNAS} and Lieb
\cite{Lieb83}.  From this work there
are three distinct functionals,
Hohenberg-Kohn (${\mathcal F}_{HK}$), Levy-Lieb ($\mathcal{F}_{LL}$),
and Lieb (${\mathcal F}_L$).  We sketch the differences.  

The HK argument established that for a non-degenerate ground state
$\Psi_0$ and its density $n_0(\br)$ from a particular $v_{ext}(\br)$,
there is a one-to-one mapping $v_{ext}(\br) \leftrightarrow n_0(\br)$ and
$E_0={\mathcal F}_{HK}[n_o] + \int d\br n_0(\br)v_{ext}(\br)$.  \textit{On
  the assumption} that \textit{every} $n(\br)$ is associated with some
$v_{ext}(\br)$ as its ground-state (``ground-state interacting
v-representability''), HK defined
${\mathcal F}_{HK}[n]:=E[v_{ext}] - \int d\br n(\br)v_{ext}(\br)$
and the associated variation principle (see
Ref. \citenum{Lieb83} at Eq. (3.10)). The difficulties with the HK
functional are (i) it is not convex on the required density domain,
(ii) it is difficult to define that domain, which is the space of
ground-state $n(\br)$, and it also is difficult to define the space of
$v_{ext}(\br)$ for which $H + V$ has a ground state, and (iii) not
every density is ground-state interacting $v$-representable. 

The Levy-Lieb functional \cite{Levy1979PNAS,Lieb83} removed both the
interacting $v$-representability problem and the non-degenerate ground
state restriction essentially by rearranging the ordinary Ritz
variational principle into variation over equivalence classes
by density, followed by variation over densities (``constrained search'' in
the DFT literature):
\bea
    {\mathcal F}_{LL}[n] &:=& \inf \Big\lbrace \frac{\langle \psi \mid H \mid \psi\rangle}%
    {\langle \psi \mid \psi\rangle} \Big\vert \psi \mapsto n\Big\rbrace \nonumber \\
    E[v] &= & \inf\lbrace {\mathcal F}_{LL}[n] + \int d\br n(\br) v_{ext}(\br) \vert %
    n \in {\mathcal I}_{N_e} \rbrace  \; .
\label{LLdefn}
\eea
Here ${\mathcal I}_{N_e}$ is the set of properly normalized densities with
finite von Weizs\"acker kinetic energy; see Ref. \citenum{Lieb83}.
The difficulty with $ {\mathcal F}_{LL}[n]$ is that it also is not convex.

Lieb then defined a convex functional 
\beq
   {\mathcal F}_L[n]:= \sup \lbrace E[v] - \int d\br v_{ext}(\br) n(\br) \rbrace
   \label{FLdefn}
\eeq
with $v_{ext}(\br)$ defined on a suitable space ($L^{3/2}+L^\infty$).
(Aside: There is a typographical error in Lieb's Theorem 3.4 (i).  It
should say ``$\tilde F(\rho)$ is not convex.'' Equivalently in our notation
``${\mathcal F}_{LL}[n]$ is not convex.'')

For most physical scientists, Lieb's functional has been regarded as
providing a firm mathematical foundation for DFT but 
a quotation from Lieb himself gives worthwhile perspective (notation
transcribed): ``Despite the hopes of HK,\ldots it is not true that 
every $n(\br)$ (even a “nice” $n(\br)$) comes from the ground state
of some single-particle potential $v_{ext}(\br)$. This problem can be
remedied by replacing the HK functional by the Legendre transform of
the energy, as is done here. However, the new theory is also not free
of difficulties, and these can be traced to the fact that the
connection between $v_{ext}(\br)$ and $n(\br)$ is extremely complicated and
poorly understood.''  40 years later some of those challenges are
unresolved.  

Regarding the one-to-one mapping, the issue of $v_{ext}(\br)$ in a
suitable space is discussed by Lieb (his Remark (ii), page 255) in terms of 
densities that do not come from some $v_{ext}(\br)$. Of specific concern are
those otherwise proper densities ``\ldots that vanish on a nonempty
open set''. If $v_{ext}$ belongs to $L^{3/2}+L^\infty$, then he asserts that
such cannot be ground state densities by the unique continuation
theorem.  But parenthetically he adds ``(Strictly speaking, this theorem
is only known to hold for $v\in L_{loc}^3$ but it is believed to hold
for $L^{3/2}+L^\infty$.)''  The argument then goes that ``if the set
of allowed $v$s can be extended properly to allow infinite $v$s, the
existence of such $n$s may not have any particular importance. The
question is very delicate, however \ldots''.  From a physical science
perspective the question also is significant because
$v \in L^{3/2}+L^\infty$ excludes the harmonic oscillator.  We return to
this below.

\subsection{Conventional discussion of functional derivatives \label{subsec:ConvFuncDeriv}}

As a practical tool for physical science, time-independent DFT almost
always relies on the existence of functional derivatives of $F[n]$ to
determine the equations used to solve for the optimizing density
$n_0$.  The existence and detailed properties of functional
derivatives remains, therefore, among the crucial open issues for
time-independent DFT.  This is the context in which, after detailed
discussion of the existence of continuous tangent functionals for
${\mathcal F}_{L}$ and ${\mathcal F}_{LL}$, Lieb posed two questions
''\ldots whose answers we cannot give but that are obviously important
for the theory''. Those two questions are about the occurrence of
continuous tangent functionals and their relationship to $v_{ext}$.
He did not discuss functional derivatives explicitly (in fact, the
term does not appear).  That was taken up by Englisch and Englisch
\cite{EnglischEnglisch1984I,EnglischEnglisch1984II}, who asserted that
on the basis of Lieb's results for continuous tangent functionals
${\mathcal F}_L$ has a proper functional derivative.

An example of how this is presented  in the DFT community
(from a treatise we respect and use) is on page 36 of
Ref. \citenum{EngelDreizlerBook}: ``In summary: The functional
derivative of $\mathcal{F}_L[n]$ exists for all ensemble
$v$-representable densities and is identical with a potential
$v_{ext}$ \ldots Moreover, for any other ``reasonable'' density $n$
\ldots one can find an ensemble $v$-representable density which is
arbitrarily close to $n$, so that the functional derivative
of ${\mathcal F }_L[n]$ again exists.''

The foregoing statement has been known not to be true since at least
2007; we defer further details to Sec \ref{sec:KnownDiffConvDFT}.
What is significant here is that, in our experience, statements such as
the foregoing are an authentic representation of the understanding of
most of the physical science segment (the majority) of the DFT
community.  Generally it is held that Levy's seminal insight
\cite{Levy1979PNAS}, Lieb's analysis \cite{Lieb83}, and Englisch and
Englisch's analysis \cite{EnglischEnglisch1984I,EnglischEnglisch1984II},
together provided a reasonable sound resolution of the limitations of
the original HK argument, save perhaps for some mathematical niceties
that are tacitly assumed to be inconsequential. Thus, restrictions on
allowable potentials are ignored for example and functional
differentiation is done routinely.

More tersely, Mermin provided the analogue to the original HK proof
for finite temperature \cite{Mermin65}.  Ensemble generalizations
\cite{Eschrig-FiniteT,FiniteTDFT2014} analogous with the conventional
ground state treatment for ${\mathcal F}_{LL}$ and ${\mathcal F}_{L}$
are obvious.  Refinements and their extension for more general systems
are described in 
Refs. \onlinecite{Lieb83,EngelDreizlerBook,YangParrBook,DreizlerGrossBook}.
Additionally, it is the conventional view that the zero temperature
results are subsumed by the appropriate limit from finite temperature
except without some of the problems of the functional derivatives that
the zero-temperature theory has \cite{Eschrig-FiniteT}.

It is safe to say that most of the attention in the DFT community is
focused on overcoming the barriers to exploiting the many advantages
to representing the original many-body problem in terms of the density
rather than the external potential. A very recent ``round-table''
paper has many details \cite{DFTroundtable}.

\subsection{ Conventional time-independent Kohn-Sham procedure \label{subsec:ConvDFT-KS}}

The variational representation in terms of a density functional
provides means for approximations that are not confined to limitations
of other quantum mechanical many-body methods. Because, however, the
proofs about ${\mathcal F}[n]$ in any of its forms provide no insight
into its structure, the strategy for exploiting the variational
representation is indirect, namely the Kohn-Sham (KS) scheme \cite{KohnSham}.
In it, $\mathcal{F}_{LL}$ is decomposed into two pieces.  Working at
$T=0$ K for simplicity, the KS strategy is to introduce an auxiliary
non-interacting system with variational functional $\mathcal{E}_s:=
T_s + E_{H} + E_{ext} + \int d\br n(\br) v_{xc}(\br)$ that has the
same number density and same external potential as the physical
interacting system. Here $T_s[n]$ is the non-interacting (KS) system
kinetic energy. The additional potential $v_{xc}$ required to maintain
the same density eventually becomes identified as connected with
exchange and correlation energies in the interacting system, hence the
subscript ``xc''.

Remark: In the conventional presentation, the variational functionals
for both the interacting and auxiliary system usually are
${\mathcal F}_{LL}$. This is expressed with admirable candor
in Ref. \citenum{DreizlerGrossBook}, p. 36: ``As a 
matter of principle, the subsequent development of the DFT formalism 
should therefore be based explicitly on the  Lieb functional.  We will 
nevertheless ignore the issue \ldots and not distinguish between the 
various flavors \ldots .''

  Introduction of the KS auxiliary system provides a definition for
the exchange energy $E_x$ in terms of a single-determinant of the KS
one-body states (``orbitals'').
Then $\mathcal{F}_{LL} + \int d\br n(\br) v_{ext}(\br)$ can be rearranged
so that the variational energy is 
\begin{equation}
  \begin{split}
    {\mathcal E}_{LLKS}[n] & := T_s[n] + E_H[n]  \\
    & + E_x[n] + E_c[n] + E_{ext}[n] 
\label{HKSLL}
  \end{split}
\end{equation}
with $E_H$ the Hartree energy, $E_c$ the DFT 
correlation energy, and $E_{ext}$ the energy from the external potential 
(usually the Coulomb interaction with nuclei or ions).

Remarks: For those conversant with variational wave-function approaches,
the KS-DFT correlation energy includes the difference
between the interacting system and non-interacting system  kinetic energies,
$T[n]$ and  $T_s[n]$.  Explicit expressions for
$E_{xc}[n]= E_x[n] + E_c[n]$ are not known except for a few special
cases. The majority of the papers that discuss ``approximate density
functionals'' concern approximations to $E_{xc}[n]$.  We return to this
below.

Variation of ${\mathcal E}_{LLKS}[n]$ with respect to the density subject to
conservation of total particle number then causes the functional
derivative $v_{xc}:=\delta E_{xc}/\delta n$ to appear as a local
potential in the
eigenvalue problem for the KS orbitals, along with the original
$v_{ext}$.  

Remarks: (a) Notice that ${\mathcal F}_{LL}$ conventionally
is assumed to have well-defined functional derivatives (an assumption
we already have highlighted).\\
(b) The Kohn-Sham decomposition is extremely helpful because
the non-interacting terms may be expressed exactly with the use of the
orbitals and because, for most systems of interest, the
exchange-correlation energy, which still must be approximated, is only
a small-magnitude correction to the non-interacting energy.

\section{Mathematical difficulties with conventional time-independent DFT \label{sec:KnownDiffConvDFT}}

%\subsection{\sout{ Conventional time-independent H-K Levy-Lieb  issues}
\subsection{Issues with conventional time-independent HK and LL functionals
  \label{subsec:KnownDiffConvHKLS}}

\textit{Issue 1} \\ The argument in Ref. \citenum{Lieb83} for the
one-to-one correspondence between density and external potential 
in the first HK theorem depends on Lieb's
\textit{assumption}, discussed above, that the unique continuation
theorem actually holds for the function space in which his argument is
formulated.  A version of unique continuation and a resulting
HK theorem was given much more recently by Garrigue
\cite{Garrigue2018} but, as we understand it, for a different function
space than used in Ref. \citenum{Lieb83}.  Hence the ground-state
``HK theorem'' that is used conventionally is not the
one for which Garrigue provided a mathematically proper foundation and
conversely. The issue of relevant spaces and corresponding conditions
upon external potentials (including unique continuation) was studied
at the same time by Lammert \cite{Lammert2018}.  He gave several
HK theorems, not just one.  Again, there are
function-space differences involved. His proofs are in Kato class
spaces, for example $K_3$.  He noted explicitly that Lieb's proofs are
in $L^{3/2}({\mathbb R}^3) +L^\infty({\mathbb R}^3)$ which is
\textit{not} in $K_3$ but then he says ``$K_3$ nearly contains the
Lieb class in some sense''.  For the thermal case, the only rigorous
proofs of the one-to-one HK theorem seem to be in a
finite-basis \cite{GiesbertzRuggenthaler2019} or on a
lattice\cite{ChayesEtAl1985}.  For explicitly finite systems, the
results for graphs are perhaps more perplexing from the conventional
viewpoint, since the first HK theorem demonstrably does
not hold \cite{PenzVanLeeuwen2021} for those systems.

\textit{Question 1.1} To what extent and in what ways are the
mathematical deficiencies of the conventional % Hohenberg-Kohn-Levy-Lieb
justification of time-independent DFT consequential for its use?  Put
another way, where and how do those deficiencies manifest themselves?
Do they matter for many physically realizable systems or are they
important only for some limited, exotic (hopefully well-defined) cases?

Remark: In the context of Lammert's remark quoted above, our question
amounts to asking about the detailed consequences of that function
space relationship. 

\textit{Question 1.2} In what way or ways must the responses to
Question 1.1 be modified for the thermal case?

\subsection{
%{Conventional time-independent KS issues}
Issues with the conventional time-independent KS procedure \label{subsec:ConvKSDFTissues}}

Briefly, at least two categories of mathematical difficulties with the
conventional KS scheme are known.  One is the $v$-representability problem:
What are the conditions under which a density $n$ can be associated
uniquely with both an interacting and a non-interacting system constructed
from it by the KS strategy? This seems to have been 
resolved (see Ref. \citenum{vanLeeuwen2003})
but we are unsure on that point. 

\textit{Issue 2} \\ The other issue has drawn much attention.  For the
Lieb functional, neither Gateaux nor Frechet functional derivatives
exist in general. It is known \cite{Lammert2007} that the supposed
proof of Gateaux differentiability in 
Refs. \citenum{EnglischEnglisch1984II} and \citenum{vanLeeuwen2003} is
incorrect. Ref. \citenum{Lammert2007} shows that differentiability can
be rescued by imposition of subsidiary conditions about the density (which
must be strictly greater than zero) \textit{and} about the first and
second derivatives of the parent ground-state wave function.  This
obviously is far from what is assumed in the conventional KS development,
\textit{vide supra}.

Remark: Even if the conditions of Ref. \citenum{Lammert2007} were to
be met, it is not obvious to us that the resulting functional
derivatives would have a straightforward operational relationship with those
derivatives commonly used in the conventional DFT development.

\textit{Question 2.1} To what extent and in what ways does the
lack of the functional derivatives that are assumed to exist
in conventional DFT manifest itself in the development and use
of approximate functionals?  
Alternatively stated,
  can a prescription or protocol be provided (such as those used for
  manipulating the Dirac delta 
function and Heaviside unit step function in the physical sciences) to account for the consequences 
of the restrictions given in Ref. \citenum{Lammert2007}, 
such that the conventional procedure could be made operationally valid?

The only other cure to the functional derivative problem so far offered
seems to be in Ref. \citenum{KvaalEtAl2014} but that requires the
use of quasi-densities.

\textit{Question 2.2} Can a detailed scheme be provided whereby
the quasi-densities of Ref. \citenum{KvaalEtAl2014} and manipulations with them
are related systematically 
to the physical densities (that are experimentally measurable
quantities) used in conventional DFT?

Remark: Many rigorous results about physical densities are known
and exploited in the development of approximate $E_{xc}$
functionals. Connecting quasi-densities systematically to those
properties is an essential pre-requisite therefore to any approach
for using quasi-densities in some reformulation of DFT.

%%%%%%%%%%%%%%%%%%%%%%%%%%%%%%%%%%%%%%%%%%%%%%%%%%%%%%%%%%%%%%%%%%%%%%%%%%%%%%%%%%%%%%%%%%%%%%%%%%%%
\section{Issues in ``orbital-free'' time-independent DFT \label{sec:OFdft}}
%%%%%%%%%%%%%%%%%%%%%%%%%%%%%%%%%%%%%%%%%%%%%%%%%%%%%%%%%%%%%%%%%%%%%%%%%%%%%%%%%%%%%%%%%%%%%%%%%%%%

We continue at  $\mathrm{T}=0$K.  
Though the ordinary KS equation has a local potential
$v_{ext}(\br) + v_{xc}(\br)$, its self-consistent solution in a basis
typically has computational costs that scale as $N_e^3$.  This scaling
is worsened by the introduction of approximations for $E_{xc}[n]$ that
depend explicitly on the KS single-particle orbitals.  Though in principle
one can find the KS potential for such an $E_{xc}$ via what is called
the optimized effective potential \cite{EngelDreizlerBook}, the computational
burden is so high that the conventional work-around is what is called
``generalized KS'' (gKS).  It amounts to taking the variational derivative of
${\mathcal E}_{LLKS}[n]$ with respect to the orbitals $\varphi_j(\br)$.  That 
worsens the computational burden because each orbital must be calculated
from a separate, orbital-dependent potential in the 
gKS equation. 
Details are irrelevant here.  To evade that
bottleneck, one may express the non-interacting kinetic energy $T_s$ as
an explicit functional of the density, obviating the use of orbitals.
Since $T_s[n]$ is not known in general as an explicit functional of $n(\br)$,
doing so requires a further approximation. This is called,
slightly misleadingly, orbital-free DFT (OFDFT). In fact, there is one
orbital $\propto n^{1/2}(\br)$.  Approximating $T_s[n]$ is
significantly more challenging
than for $E_{xc}$.  OFDFT is an area of continuing 
fundamental and practical research; see Refs. \onlinecite{WesolowskiYang2013,OFDFT2014,LKT2020}.

Thus ordinary KS calculations require accurate approximations for $E_{x}$ and
$E_{c}$, while the OFDFT form also requires an accurate approximation
for $T_s[n]$.  
The range of systems of interest is enormous, yet there are few exact
results to guide or inform development of
those required approximations.  Two routes are followed therefore.  One is
unabashedly pragmatic.  Terms in ${\mathcal E}_{LLKS}$ are
written in some physically sensible form with parameters that then are
fitted to relevant physical or chemical data (computed or
measured). Such functional approximations (called ``empirical'' in the
physics and chemistry literature) are not our priority.  Our focus is on 
the second kind, developed by imposition of whatever set of exact
properties is known as constraints.% upon the broad class of possibilities.
%Some of the commonly used constraints are noted here.
These approximations are formulated in the
framework of the KS decomposition, so we give pertinent details
next. We assume Coulombic systems.
%%%

As noted above, the  desired functional is decomposed into
the sum of a non-interacting kinetic energy, the classical Coulomb
repulsion, and the external interaction energy, denoted above as ${\mathcal E}_s[n]$, and the remainder, $E_{xc}= E_x[n] + E_c[n]$.  
%exchange-correlation contribution
%
%\begin{equation}
%  \mathcal{F}[n]=\mathcal{F}_{s}[n]+\mathcal{F}_{xc}[n].
%\end{equation}
%
%In the following, we continue to work at $\mathrm{T}=0$K for simplicity. 
Standard utilization  of the KS decomposition invokes the
explicit form of $T_s$, namely, for specified density $n$,  
\begin{equation}
  T_{s}[n]=\tfrac{1}{2}\min\limits_{\Phi \mapsto
n}\int d{\mathbf{r}}_{1}...d{\mathbf{r}}_{N_{e}}\sum_{i}^{N_{e}}|\nabla
_{i}\Phi |^{2}  \label{Tsdefn}
\end{equation}
with $\Phi $ a properly normalized single Slater determinant. (Extension to
$\mathrm{T}>0\,$K is by the corresponding 1-particle reduced density matrix.)
This leads to the commonly seen expression
\beq
T_{s}\left[\{\varphi_i[n]\}\right]  := \half\sum_{i} f_i \int \, d {\bf r} %
\:\:|\nabla \varphi_i(\mathbf{r})|^2  \; .
\label{TsDefn}
\eeq
The $f_i$ are the Fermi-Dirac occupation numbers for the KS orbitals $\varphi_i$
and the domain of integration
is $\mathbb{R}^3$ for finite systems (molecules, atoms) or the periodic volume
in 3D periodic boundary conditions.  Solving for the minimum
of $\mathcal{F}$ leads to a self-consistent eigenvalue
problem for each occupied $\varphi_i[n]$.
%For large systems
%or high temperatures, this can be an onerous computational task. 

The OFDFT utilization of the KS decomposition eschews explicit
use of the $\varphi_i$ and, instead constructs an approximate
non-interacting kinetic energy $T_{s}[n]$ (non-interacting free
energy) functional for use along with an approximate $E_{sc}[n]$.
%OFDFT therefore requires two
%approximations.
The %non-interacting kinetic
%contribution
$T_s[n]$ contribution to the total energy typically is substantially larger than the
magnitude of $E_{xc}$. Nonetheless, it is less
well-studied than the XC term, so the OFDFT approach has
several significant unsolved problems regarding the properties and behavior
of $T_s$.
%the non-interacting functionals.
%By implication, there are both large
%challenges to be overcome in reaching useful levels of accuracy and equally
%large potential gains in the practical applicability of OFDFT methodology in
%\emph{ab initio} molecular dynamics for condensed phases of very large
%bio-molecules, warm dense matter, etc.

Exact properties of various density functionals (especially XC) have
been extremely useful in the construction and improvement of practical
approximations. Thus, next we suggest several topics regarding which
mathematicians may well be able to contribute to advancing OFDFT.

\textit{Issue 3} \\ The structure and properties of $T_s[n]$ remain
somewhat obscure. In OFDFT (for convenience here still at $\mathrm{T}=0\,$K; there
is an  obvious finite $\mathrm{T}$ counterpart) there are two ways to
formulate approximations for the Kohn-Sham kinetic energy $T_s$. To embody
known properties, one-point approximations are written as
\begin{equation}  \label{1ptform}
  \begin{split}
    T_{s,1}[n]  & = T_{vW}[n]\,  \\
                & + c_{TF} \int d{\mathbf{r}}\, n^{5/3}({\mathbf{r}})  \\
                & \times f_\theta\left(n({\mathbf{r}}), \nabla n(\mathbf{r}), \nabla^2 n({\mathbf{%
r}}), \ldots \right) \;
  \end{split}
\end{equation}
with the ``enhancement factor'' $f_\theta$ to be approximated. Here $c_{TF}:=\frac{3}{10}%
(3\pi^2)^{\frac{2}{3}}$ and  $f_\theta=1$ makes that term the Thomas-Fermi kinetic
energy,
\begin{equation}
  T_{TF} :=c_{TF}\int d{\mathbf{r}} \, n^{5/3} ({\mathbf{r}}) \;.  \label{TFdefn}
\end{equation}
The first term is the von Weisz\"acker kinetic energy, 
\begin{equation}
  T_{vW} := \frac{1}{8}\int d{\mathbf{r}} \frac{|\nabla n({\mathbf{r}})|^2} {n(%
\mathbf{r})} \; .  \label{TvWdefn}
\end{equation}
The Thomas-Fermi kinetic energy is exact for a homogeneous electron gas, while the von Weizs\"acker
kinetic energy is exact for a one-electron problem or a doubly occupied two-electron problem.

Remark: There is a vast literature of
exact results on the TF problem~\cite{Teller, LS73, LS77, Lieb1981,Spruch}
 and gradient expansion corrections thereto.~\cite{K57, Hodges73, Murphy81, Salasnich07}

Two-point approximations conventionally are written as augmentations to those two limiting cases,
\begin{equation}
  \begin{split}
    T_{s,2}[n] & = T_{vW}[n] + T_{TF}[n]  \\
               & + \int\, d{\mathbf{r}} d{\mathbf{r}%
^{\prime}} \, {\mathcal{K}}[n({\mathbf{r}}),n({\mathbf{r}^{\prime}});{%
\mathbf{r}},{\mathbf{r}^{\prime}}] \; ,  \label{2point}
  \end{split}
\end{equation}
with $\mathcal{K}$ to be approximated. At least one exact result is known
about it \cite{BlancCances2005}. It also is known that the
sum of the last two terms must be $\ge 0$ because $T_{vW} \le T_s$
\cite{LevyOuYang1988}.

For reasons of computational speed 
and clarity of formulation, we have focused on one-point
approximations.  Except for the gradient expansion, some positivity
conditions on the second term in Eq.\ \eqref{1ptform}, and some properties
provable in the limit of infinite distance from a single atom, rather
little is known to guide and shape such approximations.  Practical barriers
so far have led to no higher-order spatial derivative  dependence than
$\nabla^2 n$.  

\textit{Question 3.1}:\\ For a specified highest-order $\nu$ of
spatial derivative dependence $\partial^\nu n/\partial
{\mathbf{r}}^\nu$ of a one-point approximation, is there an underlying
intrinsic property of $T_s$ that makes such an approximation
impossible in principle or subject to formidably difficult necessary
and sufficient conditions?\\
Remark: there are claims in the literature
about maximum possible order; see Ref. \onlinecite{LuoSBT2018} and
references therein.

\textit{Issue 3 -continued} \\  It long has been conjectured that one-point 
approximations cannot yield atomic shell structure because they cannot 
reproduce the interference between orbitals that arises in the
KS kinetic energy, Eq. \eqref{TsDefn}.
However, whether functionals employing higher-order derivatives, e.g.
$\nabla^2 n$, that necessarily are sensitive to small
deviations from a smooth density, could generate such structure is an
unresolved issue.

\textit{Question 3,2}: \\Is the conjecture about the non-appearance of
shell-structure in one-point approximations true and, if so, under
what conditions?\\
Remark: Observe that this question is linked to
\textit{Question 3.1}.

\textit{Question 3.3}:\\
If the answer to \textit{Question 3.2} is
negative, does obtaining shell-structure in a one-point approximation
implicate non-standard (e.g. improper signs) terms in the
gradient expansion correction to the Thomas-Fermi kinetic energy? \\
Remarks: The subsidiary matter at issue here is whether terms can be
implemented in a practical functional so as to generate the correct
shell structure in the density rather than unphysical fluctuations.
This relates to the observation, in several contexts, that the
lowest-order gradient expansion correction to the Thomas-Fermi kinetic
energy (proportional to $|\nabla n|^2$) might properly be negative,
rather than positive, as is the case in the standard gradient
expansion~\cite{K57}. The contexts include the Airy or edge
gas~\cite{LAM14}, the large-$Z$ limit of neutral atoms~\cite{CR17},
and imposition of the discontinuity in the Lindhard response function
as a constraint upon the approximation~\cite{Constantin2019}.  This
opens the possibility that shell structure could be induced because of
the reduction in $T_s$ as gradients are introduced in the density.
However, to date, approximate functionals implementing a negative
gradient expansion contribution to the kinetic energy while retaining
overall stability fail to produce shell structure in self-consistent
calculations.

\textit{Issue 3 - continued} \\
 An upper-bound to the KS kinetic energy $T_s$  that has been
  conjectured but apparently never proved \cite{Lieb1980,Lewin2020} is
\begin{equation}
  T_s \le T_{TF} + T_{vW} \; .  \label{Tsupperbound}
\end{equation}
The inequality has been invoked in constructing OFDFT $T_s$
approximations as a constraint.  It can be rationalized
\cite{TrickeyKarasievChakraborty2015} by taking the
$N_e \rightarrow\infty$ limit of the finite-system inequality due to
G{\'a}zquez and Robles \cite{GazquezRobles1982}. However, their
inequality involves a local-density approximation, hence is not an
exact result.

In the development of approximate $T_s[n]$ functionals, for
reasons of convenience, the
conjectured inequality \eqref{Tsupperbound} usually is enforced locally,
that is, as a constraint on the respective integrands in \eqref{Tsupperbound}
for an approximation to $T_s$, to wit,
\begin{equation}
  t_s({\mathbf{r}}) \le t_{TF}({\mathbf{r}}) + t_{vW}({\mathbf{r}}) \;.
\label{localconjecture}
\end{equation}
Examples of violations of local satisfaction have been discussed recently 
\cite{WittEtAl2019}.  Remark: the gauge ambiguity in using densities
such as in Eq. (\ref{localconjecture}) is well understood in the
DFT community.  

  \textit{Questions 3.4, 3.5, and 3.6}: Is the inequality \eqref{Tsupperbound}
true and, if so, under what conditions? What are the provable implications
of its use locally? Are those results different for $\mathrm{T} > 0\,$K and if
so, in what way(s)?

%%%%%%%%%%%%%%%%%%%%%%%%%%%%%%%%%%%%%%%%%%%%%%%%%%%%%%%%%%%%%%%%%%%%%%%%%%%%%%%%%%%%%%%%%%%%%%%%%%%%
\section{Additional Issues in Time-independent Density Functional Theory
  \label{sec:OtherDFTissues}}
%%%%%%%%%%%%%%%%%%%%%%%%%%%%%%%%%%%%%%%%%%%%%%%%%%%%%%%%%%%%%%%%%%%%%%%%%%%%%%%%%%%%%%%%%%%%%%%%%%%%

  It is standard practice (see, for example, Section 2.5 of
Ref.  \citenum{EngelDreizlerBook}) to generalize from DFT to spin density
functional theory (SDFT). Doing so exposes physically significant aspects
of the exchange-correlation functional that are hard to access in the original
non-spin-polarized formulation. We continue to limit discussion to
$\mathrm{T}=0\,$K.  In outline, the textbook argument goes as follows:  Define
the magnetization density $m({\mathbf r}):= n_\alpha({\mathbf r })- n_\beta(%
{\mathbf r})$  ($n_\alpha=$ ``up spin'', $n_\beta=$ ``down spin'') and an
external field  of magnitude $B({\mathbf r})$. For simplicity, take the
$\mathbf{B}$ and  magnetization directions to be aligned (though they may be
parallel or  anti-parallel).  Then to lowest-order in the field,
the KS-decomposed Levy-Lieb functional, Eq. (\ref{HKSLL}),
becomes
\begin{equation}
  \begin{split}
    {\mathcal{E}}_{LLKS}[n,m] &:= T_s[n] + E_H[n]  \\
      & + E_x[n,m] +E_c[n,m] + E_{ext}[n]  \\
    & + \mu_0\int d{\mathbf r} m({\mathbf r})B({\mathbf r})  \;.
    \label{HKSLLsdft}
  \end{split}
\end{equation}
Here $\mu_0$ is the Bohr magneton. The conventional argument is that this is
a straight-forward extension of Levy-Lieb constrained search. That is, the
usual Hamiltonian (sum of many-electron kinetic energy, electron-electron Coulomb
interaction, and external potential) is augmented by a term linear in $B$.
The KS decomposition and rearrangement gives Eq. \eqref{HKSLLsdft} and 
variation then gives spin-dependent exchange and correlation potentials in two coupled KS
equations, even for $B({\mathbf r}) \rightarrow 0$.

\textit{Issue 4} \\ However, Savin
\cite{Savin2017} has shown, by explicit example, that \eqref{HKSLLsdft} is not
bounded below for any finite $B$, so the textbook variational argument is
not valid. Earlier it had been shown \cite{CapelleVignale2001} that the SDFT
spin-dependent potentials are not unique. Though that problem can be
resolved \cite{CapelleUllrichVignale2007}, the resolution does not solve the
lower-bound issue raised by Savin. Going beyond linear order in $B$
supposedly provides the solution  but that implicates current density
functional theory \cite{TellgrenEtAl2012} and it is not obvious that it
resolves the issue of $B=0$ SDFT legitimacy.

\textit{Question 4.1}: Is there a formulation of SDFT that resolves
these issues and is connected unambiguously to the 
Levy-Lieb-Englisch formulation of the spin-independent theory?

\textit{Issue 5} \\ A fundamental problem of many-fermion physics is to
predict the existence or absence of a spectral gap for a prescribed
Hamiltonian, that is, to predict whether there is a non-zero energy
interval between the ground- and first-excited stationary states for
that Hamiltonian.  In a tour-de-force paper, Cubitt, P{\'e}rez-Garcia,
and Wolf \cite{CubittGarciaWolf2015} showed that for a certain
carefully described type of two-dimensional quantum spin system, the
spectral gap problem can be mapped to the outcome of the Turing
machine ``halting problem''. Given a well-specified input, it is
provable that the question of whether a Turing machine will halt is
undecidable.  Therefore the question of whether a system with the type
of Hamiltonian they defined is gapped or gapless also is undecidable.
Note that this is subject to the logical limitation that, as defined
by the authors \cite{CubittPerezWolf2022},
in the original proof ``gapped'' is not defined strictly as the
negation of ``gapless''.  

The direct calculation of spectral functions via DFT has not received
much attention. Ref. \citenum{JacobKurth2018} does give one
formulation with particular emphasis on degenerate states.
Viewed broadly, since DFT is, in principle, an exact
rendition of interacting many-electron quantum mechanics
and the formalism discussed by Jacob and Kurth \cite{JacobKurth2018}
is an exact result for the exact functional, a question arises.

  \textit{Question 5.1} \\ Does spectral gap undecidability apply to  exact DFT for degenerate many-electron systems, or does it fall into a different class of gapped systems?

\textit{Issue 5 continued} \\ Suppose that spectral gap undecidability
does apply for exact DFT. In practice, the theory is used with
approximations, e.g., $E_x[n]$ and $E_c [n]$ in conventional KS
calculations and, additionally, $T_s[n]$ in the orbital-free form of
the KS procedure (\textit{vide supra}). Those approximations introduce errors.

  \textit{Question 5.2} Can the errors from density functional approximations 
cause a proof of spectral gap undecidability in DFT to be nullified? For 
example, if the proof relies on halting a Turing machine, can such errors
fortuitously halt a Turing machine, thus making the evaluation of the spectral
gap decidable for degenerate states? 

  \textit{Issue 6} \\ Lieb, at the beginning of Section 4.A of Ref. 
\citenum{Lieb83}, remarked that any mathematically satisfactory definition of
the DFT variational functional ``\ldots must depend explicitly on the particle
number $N_{e}$. This fact is unavoidable and frequently overlooked.''
Various examples of such 
dependence in practice were cataloged in Ref. \onlinecite{TrickeyVela2013}.
Early investigations of approximations for $T_{s}[n]$
notably involved such dependence \cite%
{MarchWhite, Acharya, MarchParr, SearsParrDinur, GazquezRobles1982, Bartoloti, Tal}%
. Partly at least that was motivated by the possibility of generating shell
structure from $N_{e}$ dependence. The problem that explicit $N_e$-dependence could introduce, of course, is incompatibility with size consistency. (For two
systems, $A$, $B$ with ground state energies $E_{A}$, $E_{B}$, size
consistency requires that the total energy of the aggregate in the limit of
arbitrarily large separation between them be $E_{A+B}\rightarrow E_{A}+E_{B}$.) The
conventional KS form of $T_{s}$ includes the required $N_{e}$ dependence
explicitly by its sum over the kinetic energy of the occupied KS orbitals, c.f. Eq. (\ref%
{TsDefn}). Practical approximations for $E_{x}$ and $E_{c}$ either have no
explicit $N_{e}$ dependence or they pick up such dependence from use of the
KS kinetic energy density in a dimensionless indicator function \cite%
{FurnessEtAl2020}.

  \textit{Question 6.1}\\ Are there formal structures by which $N_{e}$ dependence
could be incorporated systematically in approximations of useful quality for 
$T_{s}$ (within OFDFT) and/or $E_{x}$, $E_{c}$ without violation of
size-consistency?

\textit{Issue 7} \\
Because of the differentiability issues discussed above (recall
\ref{subsec:ConvKSDFTissues}), interest recently has grown \cite{TechenkoueEtAl2019} in generating the time-independent KS potential directly by a
bijective mapping  $n(\br) \leftrightarrow v_{KS}(\br)$, 
much as is done in time-dependent DFT, \textit{vide infra}.  The
methodology is force-balance and continuity, i.e., use of the equations of motion
of the physical current density and the density respectively.
In at least one demonstration calculation \cite{RuggenthalerEtAl2202},
the potential was derived for various systems and compared with,
among others, simplified optimized effective potentials.

\textit{Question 7.1} \\
In the absence of an explicit variational density functional $E[n]$,
how is a ground-state KS potential obtained by the force-balance
scheme to be used to evaluate ground-state expectation values,
particularly the ground state energy $E_0[n_0]$? \\
Remark: Opportunism from a physical science perspective would suggest
using the $v$ from force-balance to solve for $n(\br)$ and use of that to
evaluate $E[n]$ from some approximate $E_{xc}[n]$.  The motive is
to avoid ``density-driven error'' \cite{NamEtAl20}, but the logical
difficulty is,
of course, that this procedure assumes that the inversion (bijection) is
variationally
related to the ground-state energy.

%%%%%%%%%%%%%%%%%%%%%%%%%%%%%%%%%%%%%%%%%%%%%%%%%%%%%%%%%%%%%%%%%%%%%%%%%%%%%%%%%%%%%%%%%%%%%%%%%%%%
\section{Time-dependent DFT conventional presentation \label{sec:ConvTDDFT}}
%%%%%%%%%%%%%%%%%%%%%%%%%%%%%%%%%%%%%%%%%%%%%%%%%%%%%%%%%%%%%%%%%%%%%%%%%%%%%%%%%%%%%%%%%%%%%%%%%%%%
An extension of the DFT concepts to dynamical phenomena,
%in classical and quantum systems --
time-dependent density functional theory (TDDFT), also is a very
active research area.  As just mentioned in the force-balance context,
this approach relies on the
existence of a bijective map between the time-dependent density
$n(\br,t)$ (which may take on any initial value at $t=0$) and the
time-dependent external potential $v_{ext}(\br,t)$, up to a constant.

At least from the physical scientist's perspective, proof of this
mapping is fundamentally different in character and more complicated
than the proofs for ground-state or finite-temperature DFT.  See
Ref. \onlinecite{UllrichBook} and references therein to the original
literature. One way to understand the distinction is that there is no
straight-forward variational principle for the time-dependent case.
An attempt to proceed analogously with standard procedure for the
time-dependent Schr\"odinger equation by defining an action and
varying it leads to causality problems \cite{vanLeeuwen2001}.

The focus of time-dependent DFT therefore is
almost entirely on the  $n(\br,t) \leftrightarrow v_{ext}(\br,t)$ mapping.
For pure-states, the  conventional TDDFT formulation was
put forth first by Runge and Gross \cite{RungeGross1984} with
substantial refinements by van Leeuwen \cite{vanLeeuwen1999}.  Their
result was extended to general mixed states recently by Dufty, Luo,
and Trickey \cite{DuftyLuoTrickey2018}. In all of those cases the
systems were restricted to those with densities and potentials that are analytic
in time for some domain about the initial time (an exception is the
special case of linear response).

For context about the conventional perspective, it is worthwhile to
outline the Runge-Gross \cite{RungeGross1984} proof.  It
exploits the analyticity of the potentials by considering two potentials
$v_A$, $v_B$ that differ by more than a constant.  At some order $k$, the time derivative of their
difference is non-constant (because of analyticity):
\beq
v_{A,k}(\br)-v_{B,k}(\br):=\frac{\partial^k\lbrack v_{A}(\br,t)-v_{B}(\br,t) \rbrack}{\partial t^k} \Big\vert_{t=t_0} \ne \mbox{constant}   \; .
\label{RG1}
\eeq
Then they calculate the
partial time derivative of the difference in the two quantum mechanical
currents.  
If Eq. (\ref{RG1}) is satisfied for $k=0$, the current-difference time
derivative is non-zero and the bijectivity result is trivial.  If, however,
$k > 0$, they take $k+1$ current-difference derivatives and show
that they correspond to the spatial gradient of the $k$th potential time
derivative.  By the equation of continuity, this is related to the  $(k+2)$nd 
time derivative of the density difference $n_A(\br,t)-n_B(\br,t)$ and from
there they show that this difference is non-zero provided that a certain
surface integral vanishes.  They argue that on physical grounds it must vanish.
Counterexamples and ways to deal with them have been discussed
\cite{XuRajagopal1985,DharaGhosh1987,GrossKohn1990}.

We give this rather labored summary of the Runge-Gross argument to illustrate
how very different TDDFT is, at least from the conventional physical
science perspective, from time-independent DFT.  

A quite different approach %to the proof of TDDFT
has been proposed by Ruggenthaler and van Leeuwen
\cite{RuggenthalerLeeuwen2011,Ruggenthaler2015}.  It is based on an
iterative solution to an equation relating the time-dependent density
and external time-dependent potential that follows from the
conservation laws for number density and momentum. There are two
steps: relating existence and uniqueness of solutions to that equation
to the corresponding statements of TDDFT, and proof of that existence
and uniqueness. The method does not require analyticity of the density
and external potential. While formulated for pure states it can be
extended directly to more general mixed states for both quantum and
classical descriptions.

%%%%%%%%%%%%%%%%%%%%%%%%%%%%%%%%%%%%%%%%%%%%%%%%%%%%%%%%%%%%%%%%%%%%%%%%%%%%%%%%%%%%%%%%%%%%%%%%%%%%
\section{Some Issues in Time-dependent DFT \label{sec:Issuestddft}}
%%%%%%%%%%%%%%%%%%%%%%%%%%%%%%%%%%%%%%%%%%%%%%%%%%%%%%%%%%%%%%%%%%%%%%%%%%%%%%%%%%%%%%%%%%%%%%%%%%%%

%\subsection{The TDDFT problem \label{subsec:problem}}
%%%%%%%%%%%%%%%%%%%%%%%%%%%%%%%%%%%%%%%%%%%%%%%%%%%%%%%%%%%%%%%%%%%%%%%%%%%%%%%%%%%%%%%%%%%%%%%%%%%%
  To get at the questions regarding TDDFT proofs, it is helpful to frame the problem rather generally.  Consider two
systems characterized by the Hamiltonians $H\left( t\right) $ and
$H_{1}\left( t\right) $, 
\begin{equation}
  H\left( t\right) =K+U+V(t),\hspace{0.25in}H_{1}\left( t\right)
=K+U_{1}+V_{1}(t)\;.   \label{1.1}
\end{equation}
Here, $K$ denotes the kinetic energy, $U$ and $U_{1}$ are general many-body
potentials, and $V$ and $V_{1}$ are sums of single particle potentials
\begin{equation}
  V(t)=\int d\mathbf{r}v(\mathbf{r},t)\widehat{n}(\mathbf{r}),\hspace{0.25in}%
V_{1}(t)=\int d\mathbf{r}v_{1}(\mathbf{r},t)\widehat{n}(\mathbf{r})\;. 
\label{1.2}
\end{equation}
The number density operator $\widehat{n}(\mathbf{r})$ is given by
\begin{equation}
  \widehat{n}(\mathbf{r})=\sum_{j=1}^{N}\delta\left( \mathbf{r-q}_{j}\right) \;,   \label{1.2a}
\end{equation}
where $\mathbf{q}_{j}$ is the position operator for the $j^{th}$
particle. The expectation value of some observable corresponding to an
operator $X$ is $\left\langle X\right\rangle =\textnormal{Tr}\rho X$,
$\textnormal{Tr}\rho=1.$ The trace is taken over an arbitrary complete
set of states defining the Hilbert space considered. The system state
is represented by the positive, semi-definite Hermitian operator
$\rho$ normalized to unity.  The corresponding quantities for the
second system have the same definitions but distinguished by the
subscript $1$.

The time-dependence of a state $\rho \left( t\right) $ is given by the
Liouville-von Neumann equation
\begin{equation}
  \partial _{t}\rho (t)=-i\left[ H\left( t\right) ,\rho (t)\right] ,\hspace{%
0.25in}\rho (t=0)=\rho   \label{1.4}
\end{equation}
where, without loss of generality, the initial time is taken to be $t=0$.
Accordingly, the  number densities for the two systems are
\begin{eqnarray} \label{1.5}
  n(\mathbf{r},t\mid v) &=& \textnormal{Tr}\rho \left( t\right) \widehat{n}(\mathbf{r})\equiv
\left\langle \widehat{n}(\mathbf{r});t\right\rangle, \\
  n_{1}(\mathbf{r},t\mid v_{1}) &=& \textnormal{Tr}\rho _{1}\left( t\right) \widehat{n}(\mathbf{r}%
)\equiv \left\langle \widehat{n}(\mathbf{r});t\right\rangle _{1}\;.
\end{eqnarray}
The notation $n(\mathbf{r},t\mid v)$ indicates that the density is a
space-time functional of $v(\mathbf{r},t)$, where for simplicity
we have dropped the ``ext''.  Also the subscript on the
bracket $\left\langle \widehat{n}(\mathbf{r});t\right\rangle _{1}$ indicates
an average over $\rho _{1}\left( t\right) $ with dynamics generated by
$H_{1}\left( t\right) $.  Densities $n(\mathbf{r},t)$ for which
there exists a corresponding potential $v(\mathbf{r},t)$ are
called $v$-representable and their determination is the $v$-representability
problem (\textit{vide supra}).

As stated already, the objective of
TDDFT is to show that for a given $n(\mathbf{r},t)$ the corresponding external
potential $v(\mathbf{r},t)$ is unique, i.e that there is a one-to-one mapping
of the potential and density for the given Hamiltonian $H\left(  t\right)  $.
If so, the mapping must hold as well for the Hamiltonian $H_{1}\left(  t\right)  $ with
different $U_{1},V_{1}(t)$. \textit{It follows} that for the same choice of
density there exists a unique external potential $v_{1}(t)$ such that
$n_{1}(\mathbf{r},t\mid v_{1})=n(\mathbf{r},t)$. Consequently,%
\begin{equation}
  n(\mathbf{r},t\mid v)=n_{1}(\mathbf{r},t\mid v_{1}).  \label{1.5a}
\end{equation}
This is the strongest statement of TDDFT (conditions for the initial
state are required but details are not needed for this discussion).
For the special case that system $1$ consists of
non-interacting particles, i.e., $U_{1}=0$,
this result implies that the density of an
interacting system can be reproduced by a non-interacting system with
a different external potential. That representation is referred
to as the Kohn-Sham form.

\textit{Issue 8} \\
The TDDFT proof proposed by Ruggenthaler et al. starts with an exact
representation for the functional relationship of the density and external
potential resulting from the local macroscopic conservation laws for the
density and momentum density, to wit 
\begin{equation}
  \begin{split}
  \partial _{t}^{2}n(\mathbf{r},t\mid v) =&\frac{1}{m}\partial _{i}\left[
\partial _{j}t_{ij}\left( \mathbf{r},t\mid v\right) \right. + \\
    & \left. + n({\mathbf{r}},t\mid
v)\partial _{i}v({\mathbf{r}},t)\,\right] \;.  \label{1.11}
  \end{split}
\end{equation}
(Subscripts $i$, $j$ are coordinate indices. We have left the mass
dependence explicit for clarity, though for electrons in Hartree
atomic units $m=m_e=1$.) 
This is an \textit{identity }relating the average density obtained from the
Liouville equation for the dynamics evolved under the external potential $v({%
\mathbf{r}},t)\,$. Here $t_{jk}\left(  \mathbf{r},t\mid v\right)  $ is the
average momentum flux $t_{jk}\left(  \mathbf{r},t\mid v\right)  =Tr\rho(t\mid
v)\widehat{t}_{jk}\left(  \mathbf{r}\right)$. The precise definition for the
operator $\widehat{t}_{jk}\left(  \mathbf{r}\right)  $ is known and its
average over the ensemble implies that it is a functional of
$v({\mathbf{r}},t)$.

  Now consider an arbitrary density $n(\mathbf{r},t)$ and an external
potential $w({\mathbf{r}},t)$ \textit{defined} as the solution to the
equation
\begin{equation}
  \partial _{t}^{2}n(\mathbf{r},t)=\frac{1}{m}\partial _{i}\left[ \partial
_{j}t_{ij}\left( \mathbf{r},t\mid w\right) +n({\mathbf{r}},t)\partial _{i}w({%
\mathbf{r}},t)\,\right] \;.  \label{1.12}
\end{equation}
This is an equation for $w({\mathbf{r}},t)$ in terms of the given density.
The functional $t_{ij}\left( \mathbf{r},t\mid w\right) $ is unchanged, only
its argument is different. Suppose \eqref{1.12} has a solution. Then the
Liouville equation with $w({\mathbf{r}},t)$ gives the exact conservation law corresponding to \eqref{1.11}
\begin{equation}
  \begin{split}
    \partial _{t}^{2}n(\mathbf{r},t\mid w) =&\frac{1}{m}\partial _{i}\left[
\partial _{j}t_{ij}\left( \mathbf{r},t\mid w\right) \right. + \\
    & + \left. \partial _{i}\left( n({%
\mathbf{r}},t\mid w)\partial _{i}w({\mathbf{r}},t)\,\right) \right] .
\label{1.13}
  \end{split}
\end{equation}
The difference between \eqref{1.12} and \eqref{1.13} gives the relationship
of $n(\mathbf{r},t)$ and $n(\mathbf{r},t\mid w)$
\begin{equation}
  \partial _{t}^{2}\phi (\mathbf{r},t)=\partial _{i}\left( \phi (\mathbf{r}%
,t)\partial _{i}w({\mathbf{r}},t)\right)  \label{1.14}
\end{equation}
where $\phi (\mathbf{r},t)\equiv n(\mathbf{r},t\mid w)-n(\mathbf{r},t).$ At
this point it is necessary to specify the initial conditions 
\begin{align}
n(\mathbf{r},t=0\mid w)&=n(\mathbf{r},t=0),\\
\left. \partial_{t}n(\mathbf{r},t\mid w)\right|
_{t=0}&=\left. \partial_{t}n(\mathbf{r},t)\right |_{t=0},
\end{align}
or equivalently 
\begin{equation}
\phi
(\mathbf{r},t=0)=0\;,\hspace{0.25in}\left.\partial_{t}\phi(\mathbf{r},t)\right|_{t=0} = 0.
\end{equation}
Clearly, $\phi(\mathbf{r},t)=0$ is a solution to (\ref{1.14}). However,
Ruggenthaler and van Leeuwen claim that this is the \textit{unique} solution. If
true, the proof of TDDFT follows directly as indicated below. 

\textit{Question 8.1} Is there a proof of this assertion, i.e that
$\phi (r,t)=0$ is the unique solution to \eqref{1.14}?  \\
Remark: For a
time-independent external potential, proof is straightforward
using the method of separation of variables. However, that is not the case
here. It would appear that this is a straightforward problem of classical
analysis, yet to our knowledge its proof has not yet been demonstrated.

Let us assume that the proof has been provided and consider the
consequences. At this point, the density $n(\mathbf{r},t)$ in (\ref{1.12}) is
still arbitrary. Denote the solution to (\ref{1.12}) for a given density as a
functional of that density $w({\mathbf{r}},t)\,=w({\mathbf{r}},t\mid n)\,$.
The specific functional is determined by the Hamiltonian through
$t_{ij}\left(  \mathbf{r},t\mid w\right)  $ but is otherwise universal, i.e.
$w({\mathbf{r}},t\mid n)$ delivers an arbitrary density $n(\mathbf{r},t)$ via
solution to (\ref{1.12}). Now choose that density to be the one associated
with the conservation laws for the Liouville equation with a different
external potential $u({\mathbf{r}},t)$, i.e. $n(\mathbf{r},t)\rightarrow
n(\mathbf{r},t\mid u),$ so that (\ref{1.11}) becomes
\begin{align}
\partial_{t}^{2}n(\mathbf{r},t\mid u)=&\frac{1}{m}\partial_{i}[
\partial_{j}t_{ij}\left(  \mathbf{r},t\mid u\right) \nonumber\\
 & +\partial_{i}\left(
n({\mathbf{r}},t\mid u)\partial_{i}u({\mathbf{r}},t)\,\right)  ]
\; .\label{1.20}%
\end{align}
It follows from $\phi(\mathbf{r},t)=0$ that
\begin{equation}
n(\mathbf{r},t\mid w)=n(\mathbf{r},t)=n(\mathbf{r},t\mid u) \;.\label{1.21}%
\end{equation}
Then the identity (\ref{1.11}) can be written as%
\begin{equation}
\partial_{t}^{2}n(\mathbf{r},t)=\frac{1}{m}\partial_{i}\left[  \partial
_{j}t_{ij}\left(  \mathbf{r},t\mid u\right)  +n({\mathbf{r}},t)\partial
_{i}u({\mathbf{r}},t)\,\right]  \;.\label{1.22}%
\end{equation}
Consequently, $u({\mathbf{r}},t)$ is also a solution to (\ref{1.12}). If,
as assumed, that solution is unique then $w(\mathbf{r},t)=u(\mathbf{r},t)$. Thus
the first part of TDDFT, that the potential associated with a given density is
unique, is equivalent to uniqueness of the solution to (\ref{1.12}).
Also,
since the solution provides the potential for any given density, this also
implies existence of such a potential - $v$ representability.

The preceding analysis can be repeated for a system with a different
two-particle potential $U$ $\rightarrow U_{1}$. Then the solution to
(\ref{1.12}) for system with $U_{1}$ gives an external potential for
that system for any given density. Now choose that density to be the
one for the system with $U$. The resulting unique external potential then
reproduces that density from the system with $U_{1}$. The special case
of $U_{1}=0$ gives the Kohn-Sham representation: an interacting system
density can be represented by an appropriate non-interacting system.

In summary, the solution to (\ref{1.12}) for a given density implies that
the density is $v$-representable. Furthermore, if the solution is unique, then
there is a one-to-one relationship between $n(\mathbf{r},t\mid v)$ and the
potential $v(\mathbf{r},t)$. Finally, the density from a potential
$v(\mathbf{r},t)$ and pair interaction $U$ $\ $can be generated from a
different potential $v_{1}(\mathbf{r},t)$ for a system with pair interaction
$U_{1}$. A proof of these conditions of TDDFT therefore reduces to a proof of
the existence and uniqueness of solutions to (\ref{1.12}).

\textit{Question 8.2}: Can the existence and uniqueness of
solutions to (\ref{1.12}) be proven? \\
Remark: The basic equations of the preceding discussion of TDDFT are
those of continuum mechanics and are of the same form regardless of
their basis in quantum mechanics (pure or mixed states) or in
classical mechanics. Only the explicit form for the macroscopic stress
tensor functional $t_{ij}\left( \mathbf{r},t\mid u\right) $\ differs
among the cases.

\subsection{Issues with proposed answer to Question 8.2}

\label{subsec3}

An affirmative response to the Question 8.2 has been put forth
but poses it own difficulties.

To proceed, first rewrite (\ref{1.12}) in a notation similar to
that of Ref. \onlinecite{RuggenthalerLeeuwen2011}, 
\begin{equation}
Qw({\mathbf{r}},t)\,=q\left(  \mathbf{r},t\mid w\right)  -\partial_{t}%
^{2}n(\mathbf{r},t)\label{1.23}%
\end{equation}
where $q\left(  \mathbf{r},t\mid w\right)  \equiv\frac{1}{m}\partial
_{i}\partial_{j}t_{ij}\left(  \mathbf{r},t\mid w\right)  $ and $Q$ is the
linear differential operator $Q\equiv-\frac{1}{m}\partial_{i}n({\mathbf{r}%
},t)\partial_{i}$ . Consider first the operator $Q$. For \emph{appropriate
homogeneous boundary conditions} $Q$ is self-adjoint%
\begin{equation}
\left(  \phi,Q\psi\right)  \equiv-\left(  \phi,\frac{1}{m}\partial
_{i}n({\mathbf{r}},t)\partial_{i}\psi\right)  =\left(  Q\psi,\psi\right)
\;,\label{1.25}%
\end{equation}
where the scalar product is for integrable functions over the system volume.
Furthermore, its spectrum is positive
\begin{equation}
Q\psi\equiv-\frac{1}{m}\partial_{i}n({\mathbf{r}},t)\partial_{i}\psi
=\lambda\psi\label{1.26}%
\end{equation}%
\begin{equation}
\left(  \psi,Q\psi\right)  \equiv\frac{1}{m}\left(  \partial_{i}%
\psi,n({\mathbf{r}},t)\partial_{i}\psi\right)  =\lambda\left(  \psi
,\psi\right)  \geq0 \;,\label{1.27}%
\end{equation}
for positive densities $n({\mathbf{r}},t).$ Finally, $\lambda=0$ is an
eigenvalue with eigenvector $\psi=1$ (or any constant). The right side of
(\ref{1.23}) is orthogonal to this zero eigenvector, so the Fredholm conditions
are met for inverting $Q$%
\begin{equation}
w({\mathbf{r}},t)\,=Q^{-1}\left[  q\left(  \mathbf{r},t\mid w\right)
-\partial_{t}^{2}n(\mathbf{r},t)\right]  \;.\label{1.28}%
\end{equation}

\textit{Issue 9}: \\
A constructive approach to a solution to (\ref{1.28}) can be attempted via
iteration from an initial trial solution $w^{(0)}({\mathbf{r}},t)\,$ and
successive approximations,
\begin{equation}
w^{(n+1)}({\mathbf{r}},t)\,=Q^{-1}\left[  q\left(  \mathbf{r},t\mid
w^{(n)}\right)  -\partial_{t}^{2}n(\mathbf{r},t)\right] \; ,\label{1.30}%
\end{equation}
or%
\begin{equation}
w^{(n+1)}({\mathbf{r}},t)\,=\mathcal{G}\circ w^{(n)}\left(  \mathbf{r}%
,t\right) \;.\label{1.31}%
\end{equation}
Here $\mathcal{G}\circ$ denotes the nonlinear "map"
\begin{equation}
\mathcal{G}\circ x({\mathbf{r}},t)\,=Q^{-1}\left[  q\left(  \mathbf{r},t\mid
x\right)  -\partial_{t}^{2}n(\mathbf{r},t)\right]  \;.\label{1.32}%
\end{equation}
The convergence of the sequence $\left\{  w^{(n)}\right\}  $ follows if it can
be established that $\mathcal{G}$ is a contraction mapping%
\begin{equation}
\mathcal{G}\circ\left[ w^{(n)}\left( \mathbf{r},t\right) - w^{(n-1)}\left(
\mathbf{r},t\right)  \right]  \leq\left[  w^{(n)}\left(  \mathbf{r},t\right)
-w^{(n-1)}\left(  \mathbf{r},t\right)  \right] \; ,\label{1.33}%
\end{equation}
or equivalently
\begin{equation}
w^{(n+1)}({\mathbf{r}},t)-\,w^{(n)}({\mathbf{r}},t)\leq\left[  w^{(n)}\left(
\mathbf{r},t\right)  -w^{(n-1)}\left(  \mathbf{r},t\right)  \right]
\;.\label{1.34}%
\end{equation}
Convergence, $w^{(n)}\left(  \mathbf{r},t\right)  \rightarrow w\left(
\mathbf{r},t\right)  $, implies that it is the desired solution to
(\ref{1.28}) and that it is a fixed point of $\mathcal{G}$%
\begin{equation}
w({\mathbf{r}},t)\,=\mathcal{G}\circ w\left(  \mathbf{r},t\right)
.\label{1.35}%
\end{equation}
These inequalities are measured with respect to some norm, e.g.%
\begin{align}
\left\Vert \mathcal{G}\circ \right. & \left[  w^{(n)}\left(  \mathbf{r},t\right)
  -w^{(n-1)}\left(  \mathbf{r},t\right)  \right]  \left. \right\Vert  \nonumber \\
& \leq\left\Vert
\left[  w^{(n)}\left(  \mathbf{r},t\right)  -w^{(n-1)}\left(  \mathbf{r}%
,t\right)  \right]  \right\Vert  \;.\label{1.36}%
\end{align}

\textit{Questions 9.1 and 9.2} What is an appropriate function space
for analysis of the map $\mathcal{G}$? Can suitable bounds be found to
establish that it is a contraction mapping? See Ruggenthaler et
al.\cite{Ruggenthaler2015} for the extent to which these issues have
been addressed.

%%%%%%%%%%%%%%%%%%%%%%%%%%%%%%%%%%%%%%%%%%%%%%%%%%%%%%%%%%%%%%%%%%%%%%%%%%%%%%%%%%%%%%%%%%%%%%%%%%%%
\section{Concluding remarks \label{sec:remarks}}
%%%%%%%%%%%%%%%%%%%%%%%%%%%%%%%%%%%%%%%%%%%%%%%%%%%%%%%%%%%%%%%%%%%%%%%%%%%%%%%%%%%%%%%%%%%%%%%%%%%%

A striking feature of both time-independent DFT and time-dependent DFT
is that they are formulated in terms of theorems that do not provide any
mechanical recipe (e.g.\ perturbation theory) for constructing
approximations.  Therefore, even cursory examination of the literature
on development of density functional approximations will convince one
of the influence and value that bounds, limits, asymptotics and
similar rigorously provable properties have had on such developments. What we
have delineated here is, in a way, self-serving, in that we are
certain we would be helped by having some answers to the questions posed.
Leads to other questions of rigorous nature about DFT and TDDFT may be
found in a recent ``round-table'' paper \cite{DFTroundtable}.

Our objective has been to provide a personal perspective of DFT and
its limitations in foundation and application. The discussion of the
time-independent case has been from the traditional variational
formulation and the attendant, related difficulties in
applications. This is the most common approach currently in use. The
subsequent discussion of TDDFT has been from the force balance
formulation and an elaboration of the basis for applications from a
time-dependent Kohn-Sham single particle dynamics. The literature in
both cases is extensive.  We have highlighted selected specifics for
attention. It is our hope that interested mathematicians will see
opportunities for contributions based on tools typically not familiar
to the DFT physical sciences community.
    
DFT is widely practiced across the most complex problems of 
physics, chemistry, and materials sciences, with growing
use in biomolecular areas as well. Its popularity arises 
from the capacity to formulate approximations to questions for which other
traditional many-body methods such as perturbation theory, small
parameter expansions, and simulations are limited. In application, 
typical DFT approximations 
are inherently uncontrolled in a technical sense. For example
\textit {a priori} error estimates
are not  given. But wide experience over more than a half-century
has led to ever increasing intensity of use in spite of the remaining
open problems. It is expected that the contributions sought here from
the mathematical community will mostly strengthen the confidence in
current approximations rather than to undermine or negate their
continued practice.

\begin{acknowledgments}

We are indebted to an anonymous reviewer of the original manuscript
for extensive comments that led to complete reshaping of our introductory
discussion, for informing us of some references in the mathematical
literature with which we were only dimly or not at all acquainted,
and for stimulating us thereby to revise our presentation of several
of the issues and questions.  
  
JWD, JW, and TST were supported by U.S. Department of Energy, Office
of Science, Basic Energy Sciences under Award No.
\mbox{DE-SC0002139}. SBT's work on OFDFT was
supported by that grant.  ACC and HFR were supported by U.S. National
Science Foundation grant \mbox{DMR-1912618}. SBT's work on DFT fundamentals
was supported by that grant.  AAM was supported as part
of the Center for Molecular Magnetic Quantum Materials, an Energy
Frontier Research Center funded by the U.S.  Department of Energy,
Office of Science, Basic Energy Sciences under Award
No. \mbox{DE-SC0019330}.  SBT's work on fundamental issues related to spin
and magnetic DFT was supported by that EFRC grant. 

\textit{Conflict
  of Interest} On behalf of all authors, the corresponding author
states that there is no conflict of interest. 

This paper has no associated data.

\end{acknowledgments}

 \appendix
    \section{Generic Overview of DFT \label{sec:Appendix}}

Though most of what follows is completely generic to any quantum
mechanical single-species system, our focus is on many-electron systems.
We note, for example, that there is active work on nuclear many-body DFT;
see Ref.\ \citenum{Colo2020} and references therein.

In the simplest case the Hamiltonian describes $N$
particles interacting pairwise (e.g., Coulomb interactions) subject to
an external single-particle potential that couples to each particle,
$v_{ext}(r)$. (Any possible time dependence is left implicit to simplify
the notation here.) The states of interest (wave function, density
matrix) are ``extremal" states (e.g., ground state, mechanical
equilibrium) that are fixed by conditions involving the Hamiltonian.
Hence those states are 
functionals of $v_{ext}$. Properties of interest (e.g. energy, free energy,
magnetization, etc.) are expectation
values of appropriate operators in these states, hence they inherit a
functional dependence on $v_{ext}$.  In particular the number density
$n(r)$ defined in this way is a functional of $v_{ext}$, denoted
$n(r|v_{ext})$.

For reasons of insight and computational accessibility mentioned in
the main text, it typically is preferable to express properties of
interest as functionals of $n$ rather than of $v_{ext}$. This change
of variables can be implemented if there is a one-to-one relationship
$v_{ext}(r)\leftrightarrow n(r|v_{ext})$, e.g., via a Legendre
transformation (subject to certain conditions on the functional
representing the property considered). The first task of DFT thus is
to establish this bijective relationship of the density and external
potential. Two complementary approaches have been used.

The first (historically) \cite{HohenbergKohn} is based on variational
principles showing that the density associated with a given potential
provides the extremum of a certain functional (or action in the
time-dependent case). The convexity of the functional assures the
uniqueness required. A second approach is based on the force balance
for these states, or specifically, the conservation law for the local
momentum density. Both approaches accomplish the goal formally, but
without complete mathematical rigor. Specifically, the function space
for $v_{ext}(r)$ and that for $n(r)$ have not been fully characterized
within the proof. As detailed in the main text, this deficiency
remains an open problem for all states considered: time-independent,
time-dependent, ground state, and finite temperature.
    
The variational approach also requires conditions on the associated
functional to allow functional differentiation. Existence of such
functionals in general remains an open problem, related to the above-mentioned 
characterization of function spaces. The force balance approach does
not require functional differentiation, thus avoids this difficulty.
    
Important practical problems remain after the bijectivity is
proved. The first is to know the functional $n(r|v_{ext})$, and the
second is to know the corresponding functional for the desired
property (e.g. energy, free energy, etc.) These two tasks do not
arise as separate problems 
 in the variational approach because the extremum of the
functional is identified as the primary property of interest
(e.g. ground state energy or free energy). In the force balance
approach the density can be calculated but the dependence of a 
property of interest upon that density remains to be fixed.
    
In practice, the calculation of the density for a given potential is
accomplished by a mapping of the DFT for the interacting particle
system to that for a non-interacting system, the Kohn-Sham
representation. In the variational formulation this is straightforward
once the existence of the functional derivative (or its equivalent)
can be established. Once again this can be done in the force balance
approach without need for the functional derivative.
        
%%%%%%%%%%%%%%%%%%%%%%%%%%%%%%%%%%%%%%%%%%%%%%%%%%%%%%%%%%%%%%%%%%%%%%%%%%%%%%%%%%%%%%%%%%%%%%%%%%%%

%%%%%%%%%%%%%%%%%%%%%%%%%%%%%%%%%%%%%%%%%%%%%%%%%%%%%%%%%%%%%%%%%%%%%%%%%%%%%%%%%%%%%%%%%%%%%%%%%%%%

\end{document}